\documentstyle[epsf]{mn}

\newif\ifAMStwofonts

\AMStwofontstrue

\def\kms{km~s$^{-1}$}

\def\degree{$^{\circ}$}

\def\ga{\mathrel{\hbox{\rlap{\hbox{\lower4pt\hbox{$\sim$}}}\hbox{$>$}}}}

\def\la{\mathrel{\hbox{\rlap{\hbox{\lower4pt\hbox{$\sim$}}}\hbox{$<$}}}}

\title[On the properties of HI shells in the SMC]{On the properties
of HI shells in the Small Magellanic Cloud}
\author[D.~Hatzidimitriou et al.]
       {D.~Hatzidimitriou$^1$, S.~Stanimirovic$^2$, F.~Maragoudaki$^3$,
       L.~Staveley-Smith$^4$,
\newauthor A.~Dapergolas$^5$ and E.~Bratsolis$^6$\\
$^1$Physics Department, University of Crete, P.O. Box 2208,
GR 710-03 Heraklion, Crete, Greece\\
$^2$Arecibo Observatory, NAIC/Cornell University, HC 3 Box 53995,
Arecibo, Puerto Rico 00612, USA\\
$^3$Section of Astrophysics, Astronomy and Mechanics, Department
of Physics, University of Athens, GR-15784 Athens, Greece\\
$^4$Australia Telescope National Facility, CSIRO, P.O. Box 76,
Epping, NSW 1710, Australia\\
$^5$Astronomical Institute, National Observatory of Athens, P.O.
Box 20048, GR-11810 Athens, Greece \\
 $^6$ Department Traitement du Signal
et des Images, Ecole Nationale Superieure des Telecommunications,\\
Telecom Paris, 46 rue Barrault, 75634, CEDEX 13, France\\}

\date{Accepted
      Received
      in original form 2003, November 5th}

\pagerange{\pageref{firstpage}--\pageref{lastpage}}

\pubyear{2003}

\begin{document}

\label{firstpage}

\maketitle

\begin{abstract}
\noindent There are 509 expanding neutral hydrogen shells
catalogued in the Small Magellanic Cloud (SMC), all apparently
very young, with dynamical ages of a few Myr. To examine their
relationship with young stellar objects we cross-correlated the
shell catalogue with various catalogues of OB associations, super
giants, Cepheids, WR stars, supernova remnants, and star clusters.
The incidence of chance line-ups was estimated  via Monte-Carlo
simulations, and found to be high. However, it is important that
there are 1.5 times more shells that are {\em not} spatially
correlated to an OB association, than shells that are. Moreover,
59 of the 509 shells lie mainly in low stellar density fields and
have no young stellar objects associated with them, and therefore
no obvious energy source. It is shown that, on the whole, the
properties of these ``empty" shells are very similar to the
properties of the rest of the shells, once selection biases are
taken into account. In both cases, the shell radius and expansion
velocity distribution functions are consistent with the standard
model, according to which shells are created by stellar winds and
supernova explosions, as long as all shells were created in a
single burst and with a power-law distribution of the input
mechanical luminosity. This would indicate a burst of star
formation. This interpretation, however, cannot explain why the 59
shells, with no young stellar counterparts, show almost exactly
the same behavior as shells with OB associations within their
radius. Gamma ray bursts could account for some but certainly not
for the majority of the "empty" shells. Many "empty" shells
including most of the high luminosity ones, are located in the NW
outer regions of the SMC,  and may be associated with a
chimney-like feature that is known to be exist in that area.
Finally, it is noted that turbulence is a promising mechanism for
the formation of the shell-like structures, but direct comparison
with the observations was not possible, at this stage due to lack
of detailed models.
\end{abstract}



\section{Introduction}

\label{s:intro}

High resolution neutral hydrogen (HI) maps of the Small Magellanic
Cloud (SMC) (Staveley-Smith et al. 1997, hereafter Paper I)
revealed a complex system of more than 500 `holes' surrounded by
shells of higher density. Such shells, are common in  gas-rich
galaxies, and have been catalogued in the Milky Way (e.g.
Tenorio-Tagle \& Bodenheimer 1988) as well as in several nearby
spirals and irregulars, such as the spirals M31 and M33 (Brinks \&
Bajaja 1986), the dwarf irregular galaxies  IC10 (Shostak \&
Skillman 1989, Wilcots \& Miller 1998), Holmberg II (Puche et al.
1992)and IC2574 (Walter \& Brinks 1999) and the Large Magellanic
Cloud (LMC) (Kim et al. 1999).

The origin of HI shell-like structures has been the subject of
debate for more than two decades, but the issue is far from
settled as yet.  Several different mechanisms for HI shell
formation have been proposed over the years. They can be
classified in the following two main groups.

\begin{enumerate}
\item {\em Mechanisms that require an energy source central to the
shell structure}.\\
Such mechanisms appear to provide  the natural explanation for
expanding structures with some spherical symmetry. The assumed
energy source  could be a single young massive star, or an OB
association, or a young star cluster, or some more exotic object
such as a gamma-ray burst (GRB).

In the ``standard'' picture, HI shells result from the combined
effect of hot stellar winds from O and B stars and of supernova
(SN) shocks (e.g. Weaver et al. 1977; McCray \& Kafatos 1987).
Already in the 80's it was recognized (Heiles 1984; Tenorio-Tagle
\& Bodenheimer 1988) that this explanation could not account for
all types of observed structures, especially for rapidly expanding
supershells. More recently, high resolution HI velocity maps in
Local Group galaxies have greatly revived interest in the subject.
For example, in the case of the dwarf irregular galaxy Holmberg
II, Rhode et al. (1999) failed to find the remnant stellar
populations within the shells that would be consistent with the
observed luminosity and age distribution of the shells. Although
the energy requirements adopted by these authors may have  been
significantly overestimated, and hence the limits derived for the
necessary remnant populations may be in error (as suggested by
Stewart \& Walter 2000, and through different reasoning, by
Elmegreen \& Hunter 2000),  there remains the fact that several
shells in Holmberg II are actually located in regions of very low
optical surface brightness with no indication of recent star
formation (see also Bureau \& Carignan 2002). As another recent
example, we mention the case of the LMC, where no tight
correlation was found between HI shells  and OB associations (Kim
et al. 1999), although many of the shells appeared to be
(dynamically) young enough  for the presumed OB association to be
still recognizable as such. Similarly, a spatially poor
correlation between HI shells and OB associations was found
recently in the Magellanic Bridge, a tidal bridge of gas between
the Magellanic Clouds (Muller et al. 2003). HI shells in the
Bridge also appear dynamically much younger than the corresponding
OB associations. In the Local Group spirals M31 and M33 again no
strong one-to-one correlation has been found between shells and
OB-associations, although there are specific examples of HI
structures with excellent association with sites of star formation
(see review by van der Hulst 1996).

The recent discovery of GRBs led to the realization that such
explosive phenomena occurring within a galactic disk would leave
rapidly expanding shell-like features in the interstellar medium
(ISM), which might account for some of the most luminous shells
observed (Loeb \& Perna 1998;  Efremov et al. 1999; Perna \&
Raymond 2000). Different authors have proposed different ways of
distinguishing shells  formed by a GRB or by an OB association
(see {\em Section}~\ref{s:theory}). Knowledge of the percentage of
supershells that are likely to be associated to GRBs, would lead
to important constraints on the energetics and  rates of GRBs.

\item {\em Mechanisms that do not require a central source}. \\
One such mechanism involves collisions of high velocity clouds
(HVC) with a galactic disk (Tenorio-Tagle 1981;  Tenorio-Tagle et
al. 1986), which could reproduce, under specific assumptions,
large and  energetic features, reminiscent  of some of the large
HI shell structures in our Galaxy. For example, this mechanism is
a possible explanation for the formation of the very luminous
''empty'' supershell in the Southern  Milky Way (McClure-Griffiths
et al. 2000).

The non-linear evolution of a self-gravitating disk can also lead
to formation of shell-like features, that are not related to a
central energy source (Wada \& Norman 1999; Wada {\it et al.}
2000). Earlier, Elmegreen (1997) also suggested that HI
''bubbles'' (or shells) could result naturally from the turbulent
nature of the interstellar medium, i.e.  it is possible that most
of the structure of the ISM is the result of natural gaps and
holes in the fractal gas distribution caused by turbulence;
supernovae, stellar winds, and ionizing radiation partially fill
these gaps with hot and warm ionized gas, but they need not
structure the ISM much. More recently, Dib \& Burkert (2003) used
numerical simulations to show that the large-scale turbulence
coupled with thermal effects can result in the formation of holes
and shell-like features whose sizes are compatible with
observations.

Bureau \& Carignan (2002) proposed that ram pressure of the
intergalactic medium (IGM) must be considered when studying the
large and small-scale structure of HI in a low-mass star forming
galaxy (in their case, HoII). Ram pressure (of the IGM) can create
holes in a dense gaseous disk, as well as enlarge pre-existing
holes created by supernova explosions. Unfortunately, no
simulations or detailed calculations exist to verify and quantify
this process.

Finally, Stewart \& Walter (2000) proposed that (pairs of)
supershells in spiral (massive) galaxies could  result from the
localized flaring of a pair of radio lobes formed by jets ejected
from the galactic nucleus during an active phase. This mechanism
is only mentioned here for completeness, but it obviously does not
apply to the SMC case.

\end{enumerate}

All these different processes are not necessarily mutually
exclusive and they may well all be taking place, to different
degrees of importance. Whichever physical process has resulted
into the formation of a certain shell, its subsequent evolution
and therefore its observational characteristics (such as its
radius, expansion velocity, mass, morphology) get modified by
effects that are not related to the origin of the shell. One such
effect that has been investigated in detail is radiation pressure
from field stars (Elmegreen \& Chiang 1982), which can exert an
outward force on a large shell of gas and dust in the ISM.  This
radiative force increases with increasing shell size, so a
sufficiently large shell can expand at an  ever-increasing speed
to a size of 1 kpc or more. This process could obviously attenuate
any original differences  in the dynamical properties of large,
relatively old, shells  that actually have different types of
origin. Magnetic fields may also alter significantly the shell
evolution, but they are not taken into account in current models
(e.g. van der Hulst 1996). Metal abundance is also an important
constituent of shell evolution, as it affects cooling.  Other
effects that may modify the evolution of a shell include
interactions between neighboring shells, inhomogeneity of the
ambient ISM (e.g. McClure-Griffiths et al. 2000), globular cluster
passage (Wallin et al. 1996), self-gravity, differential rotation
of the  galactic disk (e.g. Tenorio-Tagle \& Bodenheimer 1988), as
well as ram pressure of the inter-galactic medium, as also noted
above.

The rich  population of shells in the SMC provides us with an
excellent statistical sample  with which to  re-address  some of
the basic questions related to the origin and evolution of shells
and supershells in gas rich galaxies.

There are over 500 shells and supershells in the SMC (Paper I),
five times more than found in the much more  massive LMC (Kim et
al. 1999). This large number of apparently  (dynamically) young
shells -apart from being interesting per se,  probably pointing
towards a recent global burst of star formation in the SMC-
provides the ideal data  set with which to  further investigate
the origin of HI shells and supershells, in general. Shell
dynamical ages -calculated in the framework of the standard model
(see Paper I)- display a very narrow distribution with a mean age
of $5.4\pm0.1$ Myr and a standard deviation of $2.8\pm0.4$ Myr.
This would imply a highly coherent burst of star formation over
the entire main body of the SMC. This high degree of coherence, if
real, is not easily explained in the framework of stochastic
self-propagating star formation, since the mean shell age is much
smaller than the crossing time of a typical shock (see Paper I).

Assuming that the observed HI structures in the SMC are driven by
star formation, one would expect to find some correlation between
the occurrence and properties of shells and massive star formation
activity.  The search for such correlations is attempted in
Section~\ref{s:catalogues}. The statistical significance of these
correlations is also examined. In
Section~\ref{s:shell_properties}, we examine the properties of the
shells, with the emphasis on the distribution of shells on the
expansion velocity - shell radius plane. In
Section~\ref{s:theory}, we discuss the theoretical implications of
our results, for the origin and evolution of HI shells. Comparison
with different models of shell formation is also attempted.
Finally, in the last Section we present our conclusions.


\section{On the projected stellar content of the HI shells}

\label{s:catalogues}

\subsection{The HI shell catalogue}

The catalogue of HI shells and super-shells used here is to a
large extent identical to the original list published in Paper I,
which comprised of 495 giant shells and 6 super-shells. The
catalogue included information on the central radial velocity of
the shell ($V_{hel}$), the shell radius ($r_{\rm s}$), the shell
expansion velocity ($v_{\rm s}$), as well as an estimate of the
dynamical age ($T_{\rm s}$ in Myr) of a shell and the wind
luminosity required to produce the observed velocity and radius of
the shell in the framework of the standard model (see
Section~\ref{s:shell_properties}) for further details). The survey
resolution limits were 28 pc for the shell radius, and 1.7
\kms~for expansion velocity.

A re-analysis of the original data, after inclusion of short
spacings, led to the discovery of 7 additional shells and three
supergiant shells (two of these three supergiant shells were known
already but their observational properties have been revised
significantly). Properties of the 7 new shells are presented in
Table 1, while the properties of the three supergiant shells (one
new and two revised) were presented in Stanimirovic et al. (1999).
Thus, the total number of HI shells identified in the SMC has
risen to 509. The area covered by the survey was 20 square
degrees, and the spatial resolution achieved was 1.6 arcmin (or 28
pc, assuming a distance of 59 kpc).

\begin{table*}
\begin{minipage}{160mm}
\begin{center}\scriptsize
\caption{List of positions, radii, heliocentric velocities,
expansion velocities, ages and required wind luminosities for
seven new giant and supergiant shells (Stanimirovic S., PhD
Thesis, 1999).
These add to the 501 shells in Staveley-Smith et al. (1997)
(two of which have properties modified by Stanimirovic et al.
(1999)) and one new shell listed in Stanimirovic et al. (1999),
giving 509 shells altogether.}

\begin{tabular}{@{}ccccccccc@{}}
\hline
Supergiant &  RA  & DEC & \multicolumn{2}{c} {Shell}  & Heliocentric & Expansion & Age & Wind \\
Shell  & (J2000) & (J2000)  & \multicolumn{2}{c} {Radius} & Velocity   & Velocity & & Luminosity \\
  &  &  &  & r  & $V_{hel}$  & $ v_{\rm s} $ &  $T_{\rm s}$   & log$(L_{\rm s}/n_0)$ \\
  &  &  &   $(~^{'})$   & (pc)   & {\rm \kms}  &{\rm \kms}   &
                                            ($10^6$ yr)  & ($L_{\sun}$ cm$^3$) \\ \hline

34A    & 00:40:06  & $-71$:28:16  & 20  & 342 & 141  & 21 & 9.5  & 4.9 \\
84A    & 00:44:23  & $-72$:26:18  & 18  & 313 & 148 & 20 & 9.2 & 4.8 \\
182A   & 00:52:34  & $-72$:26:55  & 19  & 337 & 151 & 18 & 11.0 & 4.7 \\
198A   & 00:53:58  & $-73$:00:59  & 12  & 205 & 150 & 34 & 3.5 & 5.1 \\
389A   & 01:10:50  & $-73$:35:30  & 18  & 315 & 186 & 24 & 7.7 & 5.0 \\
394A   & 01:11:59  & $-72$:22:28  & 18  & 321 & 191 & 16 & 11.8 & 4.5 \\
411A   & 01:14:32  & $-72$:54:40  & 16  & 281 & 163 & 24 & 6.9 & 4.9 \\
\hline
\end{tabular}
\end{center}
\end{minipage}
\end{table*}

\subsection{Stellar Catalogues }
In order to identify possible young stellar counterparts of shells
and super-shells, we gathered data on OB associations and star
clusters, supergiants, Wolf-Rayet stars, Cepheids and supernova
remnants in the SMC.

\begin{enumerate}
\item {{\em OB associations and star clusters:}} Bica \& Schmitt
(1995) carried out a survey of extended objects in the SMC,
including OB associations, star clusters, and emission nebulae.
Their catalogue contains 1188 objects, of which 554 are classified
as star clusters, 343 as associations, and 291 as objects related
to emission nebulae.

\item {{\em Supergiants and Cepheids:}} We compiled a list of 362
young stars (including blue and red supergiants and Cepheid
variables) with known positions (precessed to J2000),  radial
heliocentric velocities and their accuracy and spectral types,
based on the catalogues of Maurice {\it et al.} (1989) and
Mathewson {\it et al.} (1987). Ten of the stars in this list lie
beyond the area of the HI observations (with RA greater than
$1^{\rm h} 40^{\rm m}$),  and are excluded from further
investigation. At the same time, 171 out of the 509 shells lie
beyond the area covered by the stellar list we have compiled. This
list was treated differently from the other catalogues described
in this Section, because it includes information on the radial
velocity of the stars. As will be shown later, radial velocity can
be used as a discriminator against chance line-ups.

\item {{\em Wolf-Rayet stars:}} Wolf-Rayet (WR) stars generally
have winds significantly stronger than those of their  OB stellar
progenitors.  Only 9 WR stars have been identified in the SMC, and
the coordinates were taken from the corresponding discovery papers
(Azzopardi \& Breysacher 1979; Morgan et al. 1991).

\item {{\em Supernova remnants:}} We used the catalogue of 25
known and candidate supernova remnants given by Filipovic et al.
(1998). For 12 of these supernova remnants, we used the more
accurate coordinates provided in Wang \& Wu (1992).
\end{enumerate}

Supplementary information on the spatial distribution of young
stellar populations in the SMC was derived from the analysis of
stellar populations in the SMC by Gardiner \& Hatzidimitriou
(1992), by Maragoudaki et al. (2001) and by Harris \& Zaritsky
(2004). Finally, because the completeness of the various
catalogues used was probably not uniform, and difficult to assess,
we also conducted an independent survey for OBA spectral types
based on UK Schmidt Telescope Objective Prism plates (see
following paragraph).

\subsection{Correlation Results}
The catalogue of the 509 shells and super-shells (Section 2.1) was
correlated with the catalogues described in Section 2.2 in order
to identify objects that are spatially associated with the shells,
and therefore likely to be related to their formation. Any object
with coordinates lying within the radius of a particular shell was
regarded as a possible related source. \footnote {The use of the
radius to define the spatial limit of a shell, implicitly assumes
that the shells are spherical. This assumption could only
critically affect the resulting associations for small shells.
However, small shells, particularly of Group II (see below), are
nearly circular in appearance. Therefore, our assumption is not
expected to affect significantly our results, regarding the nature
of "empty" and "non-empty" shells.} The great majority of the
shells were found to have at least one such correlation. There are
200 shells that appear to be spatially associated  with at least
one OB association, 241 shells with at least one star cluster, 71
with at least one star from the young star catalogue, 18 with at
least one WR star and 41 with at least one SNR. However, because
of the large stellar density, it is more than likely that a
significant percentage of these correlations is only due to chance
line-ups. Indeed, Monte-Carlo simulations showed that $\simeq$85\%
of spatial correlations found in this way fall into this category.
This result suggests that it is difficult to derive meaningful
conclusions about the origin of shells from spatial-only
correlations. In order to reduce the incidence of false
correlations, one can impose a radial velocity constraint as well,
retaining only correlations for which the radial velocities of
shell and object agree, for example, to within one-sigma of the
combined velocity error. Such a procedure was only possible for
the second catalogue of Section 2.2, which includes radial
velocity information.  It was found that only 21\% of the spatial
correlations found between shells and catalogue objects also
satisfied the radial velocity criterion. This low percentage
emphasizes  the high incidence of chance line-ups, also suggested
by the Monte-Carlo simulations. It becomes obvious that the high
stellar density precludes the identification of objects that could
be -with some certainty- considered to be linked with the power
source of a particular shell. The variable size of the shells
further complicates the issue, introducing additional selection
biases. Radial velocity information would provide a useful tool,
but unfortunately, the catalogue of young objects with radial
velocity values is far too incomplete. What is remarkable,
however, is that 80 shells, i.e. 16\% of the total population,
have no young objects, from any of the catalogues of Section 2.2,
lying within their radius. The case of OB associations is of
particular interest as they are often thought to be the power
source of HI shells: {\em there are 1.5 times more shells that are
{\em not} spatially correlated to an OB association  than shells
that are}.

This is the first indication that some of the shells may not be
directly related to young stellar objects, although -as it will be
seen in Section 3- they appear to be dynamically as young as the
rest of the shells. However, it can be argued that there may be
isolated young stars (of OB and A spectral types) within these
apparently "empty" shells, as the catalogues used are not
necessarily complete. In order to isolate the shells that are
indeed empty of young stars, we searched within  these 80 shells
for OB and A spectral-type stars.  The spectral classification was
performed by visual inspection (through a microscope) of  low and
medium dispersion objective prism plates, taken with the UK 1.2m
Schmidt Telescope. OB and A spectral types are very easy to
identify on these plates (Savage et al. 1985a; Savage et al.
1985b; Kontizas et al. 1988). There are two advantages in
performing this classification: (i) the observational material,
and the identification method are uniform for all shells; (ii) we
cover the  outer regions of the SMC, which were unevenly covered
by the catalogues of Section 2.2. Actually, exactly because of the
low stellar density in these areas, spectral classification is
expected to give quite accurate results (for bright stars). The
conclusion of this exercise is that 59 of the 80 previously
identified shells, are indeed empty of young stellar counterparts.
In Table 2 we provide the list of these 59 shells -along with
their properties (columns as in Table 1).

\begin{table*}
\begin{minipage}{160mm}
\begin{center}\scriptsize
\caption{List of positions, radii,heliocentric and expansion
velocities, dynamical ages and required wind luminosities for
Group II (empty) shells}
\begin{tabular}{@{}cccccccc@{}}
\hline
Shell &   RA  & DEC             & Shell  & Heliocentric &Expansion & Dynamical  & Wind \\
      & \multicolumn{2}{c} J2000 & Radius& Velocity & Velocity & Age& Luminosity \\
           &    &                        &$r_{\rm s}$& $V_{hel}$ & $ v_{\rm s} $ &
$T_{\rm s}$  & log$(L_{\rm s}/n_0)$ \\
& & & pc   & $km ~ s^{-1}$& $km ~ s^{-1}$ &$10^6 ~ yr$&$L_{\sun} ~
cm^3$ \\ \hline
11&00:34:52&-72:31:19&104.6&138.9&14.4&4.3&3.4 \\
14&00:35:14&-73:06:06& 32.6&119.5&10.2&1.9&1.95\\
17&00:35:40&-71:23:34&116.6&155.9&16.1&4.3&3.64\\
18&00:36:19&-71:19:17&53.2&144.1&6.2&5.2&1.74\\
21&00:37:13&-70:59:55&199.0&137.3&26&4.6&4.73\\
25&00:38:30&-71:41:21&116.6&136.5&14.4&4.8&3.49\\
29&00:39:12&-71:24:56&209.3&144.1&33.5&3.7&5.11\\
34&00:40:02 &-74:59:47& 48.0&135.0&2.9 &9.8&0.65\\
36&00:40:26&-71:13:13& 99.5&139.7&17.6&3.4&3.62\\
37&00:40:26&-72:58:52& 18.9&112.8&2.9&3.9&-0.15\\
40&00:40:40&-70:45:58& 99.5&139.2&14.7&4 &3.38\\
44&00:41:20&-74:19:55& 89.2&163.3&3.8&14.1&1.52\\
50&00:41:40&-70:50:50& 53.2&124.6&5.0&6.3&1.45\\
51&00:41:40&-71:07:32&  22.3&115.8&1.7&8.1&-0.71\\
54&00:41:57&-73:06:48&  29.2&169.1&2.5&6.6&0.01\\
56&00:42:07&-73:00:50&  18.9&114.0&6.7&1.7&0.93\\
58&00:42:22&-71:42:35&147.5&141.8&20.8&4.2&4.18\\
59&00:42:25&-75:15:13&221.3&152.4&10&13.2&3.57\\
60&00:42:33&-74:03:48& 34.3&113.2&5.9&3.5&1.27\\
69&00:43:14&-71:26:10& 58.3&158.1&8.1&4.3&2.14\\
71&00:43:34&-72:41:40&17.2&150.1&1.8&5.4&-0.87\\
75&00:43:40&-71:33:50&29.2&168.2&3.0&5.8&0.23\\
76&00:43:46&-72:32:41&27.4&160.6&2.9&5.4&0.15\\
78&00:44:00&-72:28:14&60.0&143.3&18.1&2.0&3.21\\
84&00:44:23&-71:29:29&33.0&131.9&4.8&4.2&0.96\\
85&00:44:27&-71:40:5& 30.9&129.1&2.9&6.2&0.25\\
89&00:44:46&-74:17:42&66.9&132.6&4.5&8.8&1.51\\
94&00:44:58&-74:25:11&37.7&134.3&2.1&10.9&0.01\\
105&00:45:58&-72:39:09&34.3&125.2&8.8&2.3&1.79\\
112&00:46:36&-74:43:18&77.2&135.9&7.6&6.0&2.3\\
119&00:47:01&-74:34:27&96.1& 96.6&11.2&5.1&3.0\\
140&00:49:05&-71:30:20&20.7&153.6& 2.2&5.6&-0.45\\
146&00:49:21&-71:40:56&56.6&154.8& 8.4&4.0& 2.16\\
155&00:50:15&-71:19:53&34.3&161.6&4.6&4.4&0.95\\
166&00:51:15&-71:23:00&49.7&160.0&6.2&4.8&1.67\\
175&00:51:34&-71:15:27&53.2&157.0&5.9&5.5&1.66\\
182&00:52:26&-71:10:58&41.2&151.0&2.9&8.4&0.53\\
211&00:54:48&-71:12:33&108.1&151.9&8.8&7.3&2.79\\
214&00:55:03&-71:06:30& 32.6&159.0&2.9&6.7&0.33\\
221&00:56:01&-72:11:25& 20.6&160.4&2.2&5.6&-0.47\\
225&00:56:24&-70:48:17& 12.0&117.7&1.5&5.1&-1.42\\
229&00:56:47&-71:39:44& 17.1&156.8&1.5&7.1&-1.14\\
234&00:57:04&-71:35:12& 46.3&156.1&3.7&7.3& 0.91\\
277&01:00:26&-71:06:10& 24.0&162.6&4.0&3.7& 0.50\\
286&01:00:49&-71:13:42& 32.6&156.4&4.0&4.9& 0.75\\
289&01:01:12&-71:10:37& 75.5&173.4&11.4&3.9&2.80\\
302&01:02:11&-71:07:26& 53.2&165.1& 5.5&5.8&1.57\\
314&01:03:50&-74:41:45& 22.3&152.9& 2.3&5.9&-0.35\\
369&01:08:53&-71:24:03& 37.7&117.7& 2.6&8.7&0.26\\
385&01:10:26&-71:12:56& 22.3&145.2& 4.2&3.3&0.48\\
422&01:15:11&-71:06:29& 15.4&175.5& 1.7&5.5&-1.07\\
441&01:17:32&-71:08:24&116.6&191.2&13.6&5.1& 3.42\\
452&01:19:04&-72:05:19& 46.3&186.0&10.9&2.5& 2.32\\
459&01:20:05&-74:35:18& 42.9&185.3& 5.4&4.6& 1.34\\
466&01:20:48&-74:40:57& 90.9&183.9& 4.6&11.9&1.80\\
473&01:21:45&-71:21:19& 25.7&206.6& 2.9& 5.1&0.08\\
476&01:22:46&-71:54:10& 70.3&189.3&12.0& 3.5&2.81\\
486&01:24:32&-74:39:05& 56.6&188.3&7.1&4.8&1.96\\
493&01:26:32&-71:53:17& 248.7&205.2&9.6&15.4&3.63\\
\hline
\end{tabular}
\end{center}
\end{minipage}
\end{table*}


\section{Shell properties}
\label{s:shell_properties}

Following the results of the previous Section, the HI shells are
classified into two groups: Group I, which contains the 450 shells
with some young stellar counterpart; Group II, which contains the
59 ``empty" shells of Table 2. The shell properties are examined
separately for Group I and Group II members.

The basic observational properties of the shells, that will be
explored, also in relation to their stellar content, include the
shell radius ($r_{\rm s}$), and expansion velocity ($v_{\rm s}$)
as described in Section 2.1, as well as morphological
characteristics of shells of particular interest (see below).
Apart from these properties, which come directly from the data
cubes without any assumptions, we also refer to two derived
quantities, namely, the dynamical age ($T_{\rm
s}=\frac{3}{5}r_{\rm s}/v_{\rm s}$ in Myr) of a shell and the
(logarithm of the) wind luminosity required to produce the
observed radius and expansion velocity ($\log L_{\rm s}/n_{\rm
o}$), where $n_{\rm o}$ is the ambient ISM density (in cm$^{-3}$).
The wind luminosity, $\log L_{\rm s}$, is only meaningful in the
framework of the "standard" model. The formulae used for the
derivation of these quantities and the inherent assumptions are
described in Paper I.

Morphologically, both groups of shells show similar general
properties: larger shells are usually not well defined and have
only partially visible rims, while smaller shells are better
defined and often show up as small donuts in a few consecutive
channels and then quickly disappear.

\subsection{Spatial distribution of Group I and Group II shells}
Figure 1 shows the spatial distribution of all 509 shells. Filled
circles mark the positions of Group I shells, Group II shells are
denoted by crosses, while the 7 most luminous Group II shells are
indicated by open squares (see Section 3.4).

\begin{figure*}
\begin{center}
\leavevmode \epsfxsize=15cm \epsffile{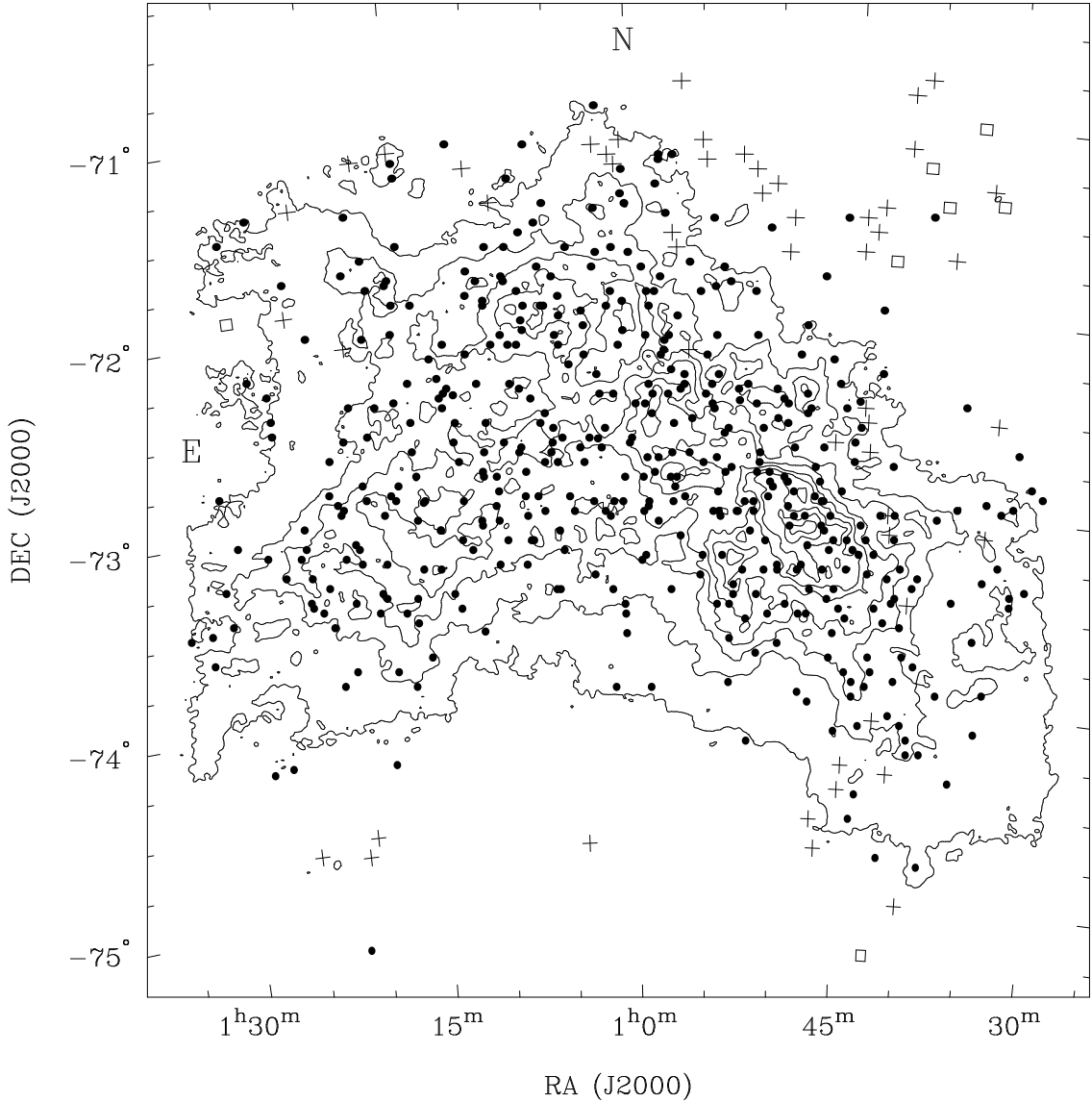}
\end{center}
\caption{Spatial distribution of the centers of HI shells in the
SMC superimposed on contours of the HI integrated column density.
The contour interval is $1.4\times10^{21}$ atoms cm$^{-2}$ and
contours at 1--10 $\times$ contour interval are shown. Filled
circles represent Group I shells while crosses represent Group II
shells. Open squares show the Group II with high luminosity, which
are discussed in Section 3.4. }
\end{figure*}

The spatial distribution of Group I shells follows the HI
distribution (indicated with contours in Figure 1), which
expectedly coincides with the distribution of young populations in
the SMC (Maragoudaki et al. 2001, Harris \& Zaritsky 2004). The
highest concentration of Group I objects is thus found along the
Bar and the Wing, with the density of objects falling
progressively in the outer regions. On the other hand, Group II
shells are mostly outliers, located beyond the second to last
contour of Figure 1. Indeed for these shells we would not expect
to find young stellar counterparts, as they are located in regions
that are known to host only intermediate and old populations (see
e.g. Gardiner \& Hatzidimitriou, 1992 and Harris \& Zaritsky,
2004).

The histograms of Figure 2 show clearly that indeed there is a
lack of Group II shells in the inner regions. This can be at least
partly explained by the fact that high stellar density would
prohibit discovery of "empty" shells in the inner, crowded
regions. On the other hand, one might interpret this difference as
indicating that the empty shells are mainly older. However, as we
shall see in paragraphs 3.1 and 3.2, there is no difference in the
dynamical ages of the two groups of shells.

It is also noteworthy that there is a significant population of
Group II shells lying in the Northwestern outer regions of the
SMC. Actually, about half the Group II shells (29) are located
within a limited area  with $0.59<RA<1.04$ and $-71.7<Dec<-70.8$,
in the NW. These NW Group II shells will be further discussed in
the following paragraphs.

\begin{figure*}
\begin{center}
\leavevmode \epsfxsize=15cm \epsffile{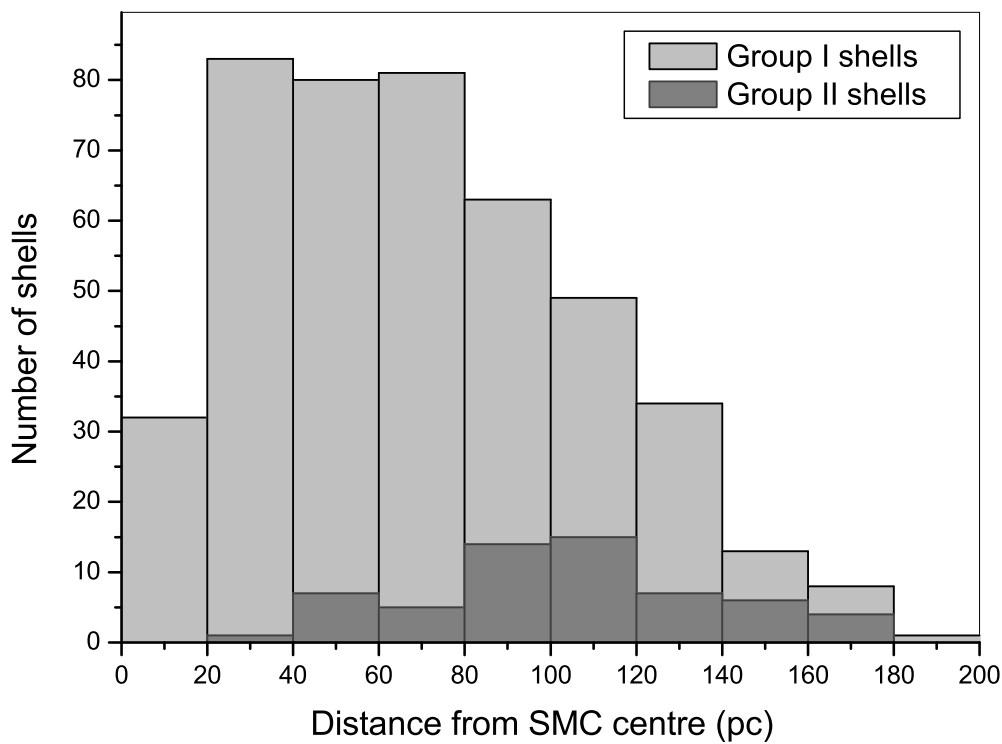}
\end{center}
\caption{Histograms of the distances from the SMC centre for Group
I (light grey) and Group II (dark grey) shells }
\end{figure*}

\subsection{Differential  shell radius and expansion velocity
distribution functions}

\begin{figure}
\begin{center}
\leavevmode \epsfxsize=9cm \epsffile{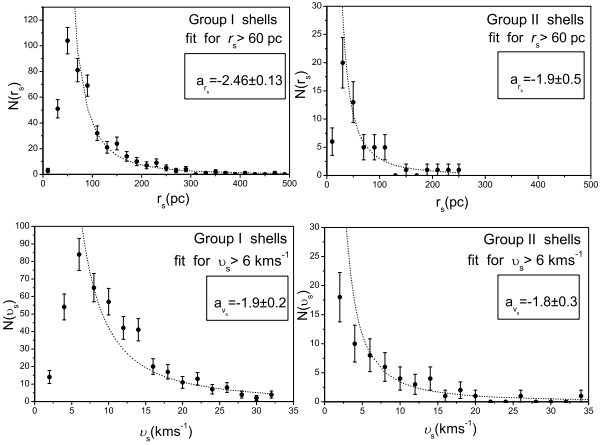}
\end{center}
\caption{The differential distribution functions of shell radius
and expansion velocity for the Group I (left panels) and for Group
II shells (right panels). Statistical errors, $N^{1/2}$ are shown.
The weighted power-law fits, $N(r_{\rm s}) \propto r_{\rm
s}^{\alpha_{\rm r_{\rm s}}}$ and $N(v_{\rm s}) \propto v_{\rm
s}^{\alpha_{\rm v_{\rm s}}}$ are overlaid. All slopes are listed
in Table 3.}
\end{figure}

The shell radii, $r_{\rm s}$, range from close to the resolution
limit of the survey to 473 pc, with a mean value of 93pc. The
expansion velocities, $v_{\rm s}$, range from close to the
resolution limit to $\simeq$ 33.5 \kms, with a mean value of 10.3
\kms. Thus, in many cases $v_{\rm s}$ is a significant fraction of
the SMC escape velocity of $\simeq$ 50 \kms. This suggests that
the ISM of the SMC is suffering from severe disruption, a result
that is also indicated by the fact that the total wind energy of
the catalogued shells deposited in the ISM is a significant
fraction ($\simeq$ 5\%) of the total HI$+$He binding energy in the
SMC (Paper I).

Figure 3 shows the differential distribution functions of the
shell radii, $N(r_{\rm s})$, and expansion velocity, $N(v_{\rm
s})$, for Group I and empty Group II shells\footnote{Although we
do not show here the differential distribution functions for all
shells (Group I and Group II together), these are almost identical
to the Group I plots and very similar to those published by Oey
and Clarke (1997), as would be expected since the samples are
almost identical.}. The shell radius and expansion velocity
distributions peaks are given in Table 3. Group I and Group II
distributions peak at significantly different values, with Group
II shells appearing in general smaller and with lower expansion
velocities than Group I shells. More than two thirds of empty
shells have $r_{\rm s}<100$ pc and $v_{\rm s}<10$ \kms. It must be
emphasized however, that chance line-ups increase with increasing
shell size. Therefore, the lack of larger, faster expanding shells
of Group II can be at least partly attributed to selection biases.
On the other hand the relative paucity of small non-empty (Group
I) shells appears to be real.

Over much of the range, $N(r_{\rm s})$ and $N(v_{\rm s})$ can be
fitted by a power-law, $N(r_{\rm s}) \propto r_{\rm
s}^{\alpha_{\rm r_{\rm s}}}$ and $N(v_{\rm s}) \propto v_{\rm
s}^{\alpha_{\rm v_{\rm s}}}$. The fit in each case has been
statistically weighed (by the inverse of the root-N errors shown
on Figure 3). The derived exponents and the corresponding error
bars  are presented in Table 3. The values of $\alpha_{\rm r_{\rm
s}}$ depend on the bin size used to derive the distribution
functions of Fig.3. Expectedly, varying the bin size leads to
different exponents, which, however, are in agreement with each
other within the quoted errors. On the other hand, the derived
exponents also depend quite sensitively on the choice of the
minimum shell radius and expansion velocity ($\rm r_{\rm s,min}$
and $\rm v_{\rm s,min}$) for the data used for fitting. This
difference is at least partly due to the different level of
completeness of the bins for small shells, but it could also be
caused by deviations from the derived single exponent law, for
smaller shells. The resolution limit of the survey is 28pc.
Considering shells larger than twice this limit (i.e. larger than
60pc), we obtain an exponent of $\alpha_{\rm r_{\rm
s}}=-2.46\pm0.13$ (for Group I shells), while, considering shells
larger than 100pc (as done by OC97), which is more than three
times the resolution limit, we get $\alpha_{\rm r_{\rm
s}}=-2.7\pm0.2$ (for Group I shells), in agreement with the result
of OC97. \footnote{This result, is identical to the result of
OC97, as would be expected, however the quoted error is
significantly smaller in our case. This is due to the fact that we
quote standard errors, while OC97 quoted rms values).} For the
expansion velocity distribution functions, we use two different
values of $\rm v_{\rm s, min}$, at 6 and 8 kms$^{-1}$, in
accordance with the limits in shell radius \footnote{The expansion
velocity limits have been chosen so that the point [$\rm r_{\rm s,
min},\rm v_{\rm s, min}$] lies close to the mean line defined by
the data points on the r-v plane. This was considered to be
necessary, since the resulting exponents for the radius and
velocity distributions are subsequently inter-compared}. For $\rm
v_{\rm s, min}=6$ kms$^{-1}$, the exponent of the power-law is
$\alpha_{\rm v_{\rm s}}=-1.8\pm0.2$ , while it becomes
$-2.1\pm0.2$ for $\rm v_{\rm s, min}=8$kms$^{-1}$. All these
values refer to Group I shells.

The same procedure is followed for Group II shells (Table 3).

The least-squares power-law fits that are shown in Figure 3, are
those derived with $\rm r_{\rm s, min}=60$pc and $\rm v_{\rm s,
min}=6$kms$^{-1}$.

As it can be seen in Table 3, all slopes are between -1.8 and
-2.9. Slopes derived using different lower limits are in all cases
in good agreement with each other, within the combined errors.
Also, $\alpha_{\rm r_{\rm s}}\simeq \alpha_{\rm v_{\rm s}}$
(within the combined errors) for all cases.  Group I and Group II
 shell distributions can be described by very similar power law fits
 (for the adopted limiting radii and expansion velocities).

\begin{table*}
\begin{minipage}{\linewidth}
\caption{\label{t:stats} The number of shells, the peak of the
shell radii distribution (for the entire sample of the particular
shell type), the lower limit adopted for the power-law fit , and
the resulting exponent of the power law (followed by the standard
error), followed by the same parameters for the shell expansion
velocity distribution. }
\begin{tabular}{lccclccc}
\hline Shell & $N_{\rm shells}$ &\multicolumn{3}{c}{Radius} &
\multicolumn{3}{c}
{Expansion velocity} \\

 Type  &    & Peak & Lower Limit& Slope & Peak & Lower Limit& Slope \\
       &    & (pc) & $\rm r_{\rm s, min}$(pc)& $\alpha_{\rm r_{\rm s}}$ & (kms$^{-1}$) &$\rm v_{\rm s, min}$(kms$^{-1}$)
       &$\alpha_{\rm v_{\rm s}}$ \\
 \hline
All shells & 509 & $\simeq$60     &   60      & $-2.39\pm0.12$ & $\simeq$8& 6 & $-2.0\pm0.2$\\
All shells &     &        &  100      & $-2.7\pm0.2$   &  & 8 & $-2.4\pm0.2$\\

Group I    & 450 & $\simeq$60     &   60     & $-2.46\pm0.13$ & $\simeq$8& 6 & $-1.9\pm$0.2 \\
Group I    &     &        &  100      & $-2.8\pm0.2$  &  & 8 & $-2.4\pm$0.2 \\

Group II   &  59 & $\simeq$30     &   60     & $-1.89\pm0.45$   &$\simeq$ 3& 6 & $-1.8\pm$0.3 \\
Group II   &     &        &  100      & $-2.9\pm1.4$  &  & 8 & $-2.3\pm$0.3 \\

\hline
\end{tabular}
\end{minipage}
\end{table*}

\subsection{Distribution on the $\log v_{\rm s}$ -- $\log r_{\rm
s}$ plane}
\label{s:v-r-plane}

Figure 4 shows the locus of Group I and Group II shells on the
$\log v_{\rm s}$ vs $\log r_{\rm s}$ plane. Both groups of shells
occupy a similar area on the $\log v_{\rm s}$ vs $\log r_{\rm s}$
plane, although most small shells belong to Group II, as noted
earlier. The solid line is the least absolute deviation fit to all
Group I shells, having a slope of $0.68\pm0.03$. The dashed line
is the corresponding line for the Group II shells, and has a slope
of $0.87\pm0.08$. The slopes derived for the two Groups of shells
differ by about twice the combined errors. For all 509 shells, the
corresponding slope is equal to $0.71\pm0.03$, i.e. very similar
to the Group I slope.

The scatter around the regression lines shown in Fig.4 is quite
large especially for Group I shells, at intermediate radii. It
would appear that most small shells lie below the regression lines
(regardless of group), pointing towards a different slope for the
smaller shells. Indeed, if we consider only shells with $\log
r_{\rm s}<1.6$ (of both groups), the slope of the corresponding
regression line approaches unity ($0.95\pm0.09$, indicating a
linear relationship between $v_{\rm s}$ and $r_{\rm s}$. For
larger shells ($\log r_{\rm s}\geq 01.6$), on the other hand, the
slope appears to be lower at ($0.58\pm0.05$). This difference is
statistically significant at the 3-sigma level and it may account
at least partly for the difference between the slopes for Group I
and Group II, since the proportion of small shells is larger
within the Group II objects. Indeed, if we only take into account
shells with $\log r_{\rm s}>1.6$, the slopes of the two groups
agree within one-sigma.

The implications of these results will be further discussed in the
following sections.

\begin{figure}
\begin{center}
\leavevmode \epsfxsize=9cm \epsffile{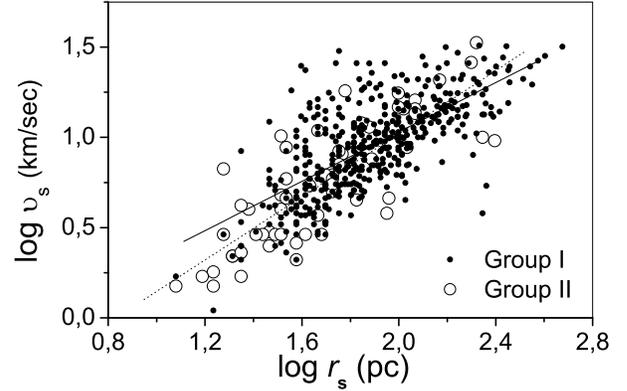}
\end{center}
\caption{The logarithm of expansion velocity (in km s$^{-1}$)
versus the logarithm of shell radius (in pc) for the entire sample
of HI shells. Filled circles mark Group I ("non-empty") shells,
and open circles mark Group II ("empty") shells. The solid line is
the least absolute deviation fit to Group I shells, while the
dashed line to Group II shells.}
\end{figure}

\subsubsection{Effect of distance from the SMC center}
 We now examine whether the distribution of Group I and Group II
shells on the $\log v_{\rm s}$ vs $\log r_{\rm s}$ plane depends
on the projected distance of a shell from the kinematic center of
the SMC (at RA 00$^{\rm h}$ 51$^{\rm m}$, Dec $-73$\degree, 1950).
We divided each group into two subgroups, the "inner" and "outer".
The former includes shells at (projected) distances smaller than
50 arcmin (i.e. $\simeq 0.85$ kpc) from the SMC center and the
latter at distances larger than 1.5 degrees (i.e. $\simeq 1.55$
kpc). Figure 5 shows the differences in the distribution of the
two subgroups on the $\log v_{\rm s}$ vs $\log r_{\rm s}$ plane,
for Group I and Group II shells (top and bottom panel
respectively).

A first examination of these figures would imply that there is a
lack of large shells (of either Group), with  $\log r_{\rm s}\geq
2.2$pc), in the inner regions of the SMC (open symbols). However,
after correcting for the different area fractions of the galaxy
for the inner and outer regions, we find {\em no} statistically
significant tendency of these larger shells to be located at
larger galactocentric distances in the SMC.

\begin{figure}
\begin{center}

\leavevmode \epsfxsize=9cm \epsffile{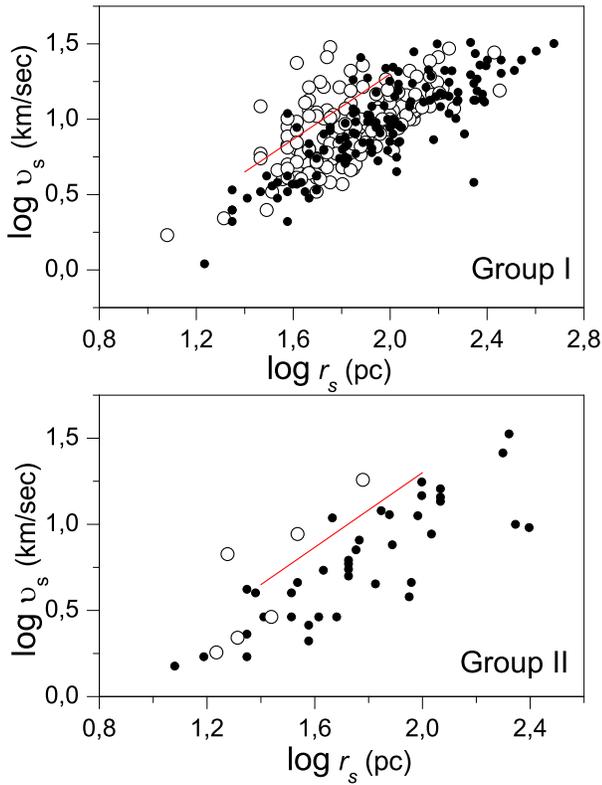}
\end{center}
\caption{The logarithm of expansion velocity (in km s$^{-1}$)
versus the logarithm of shell radius (in pc) for Group I shells
(top panel) and for Group II shells (bottom panel). Filled circles
mark shells that lie beyond 1.5 degrees in projection from the
center of the SMC. Open symbols mark shells closer than 50 arcmin
from the center of the SMC (in projection).}
\end{figure}

An interesting feature of Figure 5  is that, in the inner regions
(open circles),  there is a significant number of shells at
intermediate radii, that show a large deviation towards higher
expansion velocities, compared to the mean locus of the rest of
the shells (i.e. above the line shown on the figure, to guide the
eye). Quantitatively, the ratio of the number of group I inner
shells (open circles) that lie above the marked line to the number
of group I inner shells that lie below the line is $0.21\pm0.05$,
while the same ratio for the outer group I shells is significantly
lower, at $0.06\pm0.02$. For Group II shells, the picture is the
same, but the number statistics are much poorer (the first ratio
is $1\pm0.08$ and the second one $0.07\pm0.05$).

In the framework of the central source scenario, this surplus of
faster expanding inner shells can be interpreted as a difference
in the age distribution of shells, with the inner shells having a
significant younger component, which would be expected since the
inner parts of the SMC generally host younger populations
(Maragoudaki et al. 2001). This option can only be considered for
Group I shells. Viewed in a different way, the observed difference
may indicate that for the same expansion velocity, shells in the
inner regions tend to be smaller than the outlying ones, which
could in principle result from the difference in the ambient
density  in the inner and outer regions (Tenorio-Tagle \&
Bodenheimer 1988), being higher in the former case. The fact that
both Group I and Group II shells show very similar behavior
(although the number statistics are very poor for Group II shells)
in this respect, may indicate that the ambient density rather than
age is the decisive factor, as it would seem meaningless to
consider the ages of the central sources, if the central source
scenario is not a valid option, at least for the Group II shells.

\subsection{Search for evidence for cloud-disk collisions}
As mentioned in the Introduction, some shell-like structures can
result from the interaction of a high velocity cloud with the
gaseous disk. Simulations show (Tenorio-Tagle et al. 1986; Rand \&
Stone 1996) that important morphological differences should be
expected, between shells caused by stellar winds/SN explosions and
cloud-disk collisions.  While stellar wind/SN explosions shells
evolve isotropically, shells caused by cloud-disk collisions grow
preferentially in the direction of impact. Also, in the case of
cloud-disk collisions, there is an absence of gas along the path
of the cloud. In the case of M101 (van der Hulst \& Sancisi 1988)
some evidence exists for the  high velocity debris gas from an
ancient interaction with a high velocity cloud. Therefore, it is
relevant to investigate the morphological properties of Group II
shells. However, close inspection of the data cubes for all 59
Group II shells did not reveal any HI features that might indicate
past interaction with a high velocity cloud.

\subsection{High Luminosity Shells}
Group I and Group II shells have similar luminosities, lying
within the range $\log L_{\rm s}/n_{0} \simeq -1.5$ to $5.8$
(solar units) for Group I and between $-1.4$ and $5.1$ for Group
II shells. One might expect to find more high luminosity shells
associated with OB associations (i.e. of Group I). However, that
is not the case. The proportion of Group II high luminosity shells
(with $\log L_{\rm s}/n_{0} \geq3.5$) is the same (i.e.
$\simeq12$\%) as the proportion of Group I high luminosity shells.

We investigate here the spatial distribution of the most luminous
shells, i.e. of those with $\log L_{\rm s}/n_{0} \geq3.5$), of
either Group. For Group I, these shells are evenly distributed in
the main body of the SMC, but avoiding entirely the Wing region.
The very luminous Group II shells are numbers 17, 21, 29, 36, 58,
59, 493 (shown as open squares on Figure 1). The first five of
them are located close to each other in a region of the order of 1
degree in (projected) diameter in the North-Western (NW) outer
regions of the SMC, where as was discussed in paragraph 3.1 there
is also a large concentration of empty shells of all luminosities.
This remarkable spatial coincidence may provide an important clue
regarding the possible origin of these shells. From the remaining
two shells, no. 493 is located in the NE outer regions, while no.
59 is one of the most remote South-Eastern shells.

The five high luminosity NW Group II shells appear to be,
spatially and in radial velocity, associated with a chimney-like
feature, located at the NW side. This feature has been described
in Stanimirovic et al. (1999) as consisting of several aligned
shells and filamentary structures. Interestingly, the `chimney'
starts its propagation in the direction almost perpendicular to
the major kinematic axis of the SMC and turns to the North-East
later on. It is $\sim$ 550 pc wide. This reminds of Galactic
chimneys, collimated HI structures thought to result from
super-bubbles bursting out of the Galactic disk (Tomisaka \&
Ikeuchi 1987; Heiles  et al. 1996). Even more interesting are
several cometary-like clouds which appear to be associated with
this NW feature in the position-velocity diagrams (as an example
see Figure 6). The morphology of these clouds is similar to that
of an atomic cloud associated with the W4 chimney (Taylor et al.
1999). This cloud, $>40$ pc in size, was seen in $^{12}CO$, 21cm
line and continuum emission by Heyer et al. (1996), and is thought
to be a result of photo-dissociation of molecular material by
strong stellar winds from the stellar cluster that created the W4
chimney. The clouds that we find to be associated with the
chimney-like feature in the SMC are larger, with a typical size of
150--300 pc.

To summarize, it is possible that the five very luminous NW Group
II shells are associated with an ancient chimney in the SMC, and
therefore they may be of different origin than other Group II
shells. The question then rises, whether the rest of the Group II
shells in the same NW region could also be associated with the
same chimney.

Although this possible connection to an ancient chimney for
several Group II shells seems to be an attractive alternative, it
is by no means clear that it is a plausible explanation which can
account for all of the shell properties. We note here that all of
the 29 Group II shells that lie in this NW region are located
right in the middle of the distribution of the rest of the shells
of either group on the $\log v_{\rm s}$ vs $\log r_{\rm s}$ plane,
actually following an even tighter relationship than the rest of
Group II (or Group I) shells.

\begin{figure}
\begin{center}
\leavevmode \epsfxsize=8cm \epsfbox[25 380 580 770]{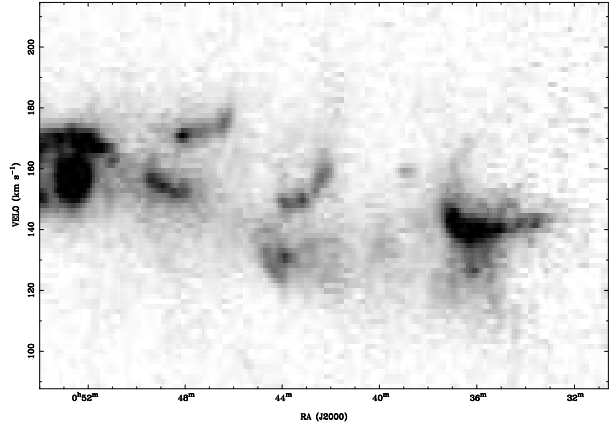}
\end{center}
\caption{A position-velocity image of the SMC at Dec $-71$\degree
$27' 11''$ (J2000), showing cometary clouds possibly associated
with the chimney-like feature located around RA 00$^{\rm h}$
40$^{\rm m}$.}
\end{figure}

\subsection{Summary of properties of Group I and Group II shells}

\begin{enumerate}

\item 12\% of all expanding shells identified in the SMC, are of
Group II, i.e. they do not appear to have an OB association or any
other young stellar counterpart within their radius. Actually,
many of them lie in regions where there are no young populations
according to optical studies.

\item Group II shells lie preferentially in the outer regions.
Part of this tendency may be caused by the fact that, in the inner
regions, chance line-ups are more likely to mask the presence of
an "empty" shell.

\item Almost half of the Group II shells lie in the outer NW area
of the SMC. There are indications that some of them could be
connected to an old HI "chimney" structure.

\item There is no convincing evidence that Group I and Group II
shells differ morphologically in any systematic and significant
way.

\item Shell radius and expansion velocity differential
distribution functions can be described by very similar power law
fits, with $\alpha_{\rm r_{\rm s}}\simeq \alpha_{\rm v_{\rm s}}$.
There are no significant differences between the values derived
for the two different shell-groups, at least for the adopted lower
limits of $\rm r_{\rm s}$ and $\rm v_{\rm s}$ used for the fits.

\item The distribution of both groups of shells on the $\log
v_{\rm s}$ vs $\log r_{\rm s}$ plane is similar, with $v_{\rm s}
\propto r_{\rm s}^{0.68\pm0.03}$ for Group I and $v_{\rm s}
\propto r_{\rm s}^{0.87\pm0.08}$ for Group II shells.

\end{enumerate}

\section{Theoretical implications}
\label{s:theory}

\subsection{Standard model for evolution of shells}
\label{s:ob-stats}

As mentioned briefly in the Introduction,   the standard model for
the evolution of giant shells (Weaver et al. 1977; McCray \&
Kafatos 1987) assumes that shell expansion is powered by stellar
winds and, especially, by SNe from the parent OB association.  It
is also assumed that the input mechanical luminosity ($L_{\rm
s}$), dominated by SNe, remains constant with time. One way to
compare properties of shells predicted by the standard model with
observations is by deriving the differential shell-size or
expansion velocity distribution functions and comparing them with
the functions inferred from observations.

There are two main ingredients when trying to predict distribution
functions $N(r_{\rm s})$ and $N(v_{\rm s})$: the spectrum of the
input mechanical luminosity function (MLF), which is often
considered either as a constant value or a power-law function
($\Phi(L_{\rm s}) \propto L_{\rm s}^{-\beta_{\rm ob}}$), and the
nature of shell creation, which could be either continuous (the
case when shells are generated continuously at a constant rate),
or a single burst (the case when all shells were created in an
instantaneous burst). The slope $\beta_{\rm ob}$ of the MLF is
related to the slope $a$ of the HII luminosity function (HII LF),
by $\beta_{\rm ob} \leq a$. Hence, as $a$ can be obtained from
observations, it usually serves as an upper limit of $\beta_{\rm
ob}$.  However, empirical measurements imply that one can easily
assume $\beta_{\rm ob}\simeq a$ in most cases (Oey \& Clarke
1997). For the various combinations of the MLF function and the
type of shell creation, one can derive the shell-size and
expansion velocity distribution functions and compare them against
observations (Fig.3, Table 3). In addition, the tight correlation
between $r_{\rm s}$ and $v_{\rm s}$ (Section 3.3) prompted us to
investigate the mean expansion velocity of an ensemble of shells,
being at different stages of their evolution, ($\overline{v}$) as
a function of shell size. We define $\overline{v}$ as:
\begin{equation}
\overline{v}(r_{0}) = \frac{\sum_{r=r_{0}} v_{i}}{N(r_{0})}.
\end{equation}

\begin{figure}
\begin{center}
\leavevmode  \epsfxsize=9cm  \epsffile{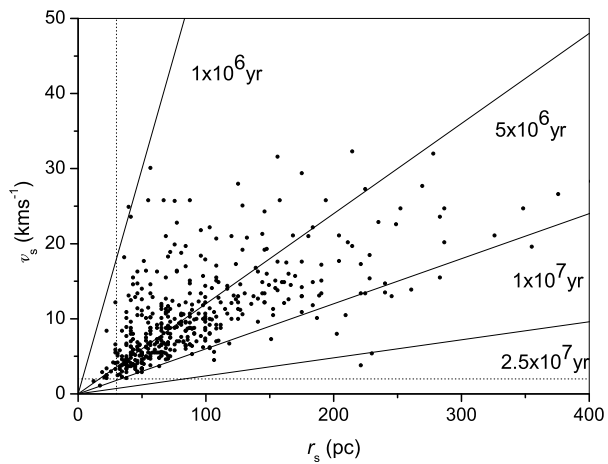}
\end{center}
\caption{\label{f:VR} Expansion velocity versus shell radius
diagram for HI shells that are found to have OB associations
within their radius.  The solid lines represent dynamical ages of
1, 5, 10 and 25 Myr. The horizontal and vertical dotted lines mark
the resolution limits of the survey. }
\end{figure}

Oey \& Clarke (1997) derived $N(r_{\rm s})$ for three different
cases within the standard adiabatic model. They considered growing
shells, as well as stalled shells. The major result of their work
was that $N(r_{\rm s}) \propto r_{\rm s}^{1-2\beta}$ is a robust
description of the shell size distribution function for both
continuous and instantaneous shell formation, for shells smaller
than $\simeq1300pc$. These shells are dominated by stalled shells.
Shells larger than $\simeq1300pc$, which are mainly growing
shells, obey $N(r_{\rm s})\propto r_{\rm s}^{4-5\beta}$.   Oey \&
Clarke (1997) compared their predictions with observational data,
particularly focusing on the SMC shell population (based on
Staveley-Smith et al. 1997) as this was the most reliable and
complete shell catalogue at the time. They found excellent
agreement with the shell size distribution predicted by the
theory, for the case of continuous formation and a power-law
luminosity spectrum. In this scenario the shell size distribution
was however dominated by stalled shells, while the shell catalogue
they used (that of Staveley-Smith et al. 1997) only included
growing shells: The latter authors did not catalogue stalled
shells, but noted that there may be $\sim50$ such shells in the
SMC, only 10\% of the total. Actually, {\it all shells} from the
HI catalogue discussed in Section 2.1 have well defined expansion
velocities. In addition, their dynamical ages are remarkably
similar. Although mechanisms like shell re-acceleration may occur
and result in shell ages being underestimated, it is hard to
imagine that this (and/or other) process will work in accord on
all shells to make the remarkably tight dynamical age
distribution. It is possible that this points to a coherent burst
of star formation that happened in the SMC. For these reasons, we
repeat here calculations by Oey \& Clarke (1997) but we only
consider growing shells based on observational constraints. In
addition, we derive the expansion velocity distribution and the
mean expansion velocity for an ensemble of shells, which were not
derived by Oey \& Clarke.

\subsubsection{Analytic expressions and comparison with observations}

We start with equations for the growth of shells in a uniform ISM
with a constant ambient density during the energy-conservation
phase: $r_{\rm s} \propto L_{\rm s}^{\gamma} t^{\alpha}$  and
$v_{\rm s} \propto L_{\rm s}^{\gamma} t^{\alpha-1}$ (MacLow \&
McCray 1988), where $\gamma =1/5$ and $\alpha=3/5$. Similarly to
Oey \& Clarke (1997) but considering {\it only} growing shells we
derive $N(r_{\rm s})$,  as well as $N(v_{\rm s})$ and
$\overline{v_{\rm s}}(r_{\rm s})$, in three different cases within
the standard adiabatic model. In all three cases power-law
behavior for all three functions can be expected.

\begin{enumerate}
\item {\bf Continuous creation, single luminosity.} Assuming that
shells are continuously created, then:
\begin{equation}
\frac{dN}{dt} = const. = \psi .
\end{equation}
If the input mechanical luminosity is the same for all shells,
then:
\begin{equation}
\Phi(L_{\rm s})=const. = L.
\end{equation}
As
\begin{equation}
N(r_{s}) dr_{s} = N(t) dt = \psi dt,
\end{equation}
then substituting $dr_{s}/dt$ we get:
\begin{equation}
N(r_{\rm s}) \propto r_{\rm s}^{1/\alpha-1}.
\end{equation}

\noindent In a similar manner, starting with:
\begin{equation}
N(v_{s}) dv_{s} = N(t) dt = \psi dt,
\end{equation}
and substituting $dv_{s}/dt \propto t^{\alpha -2}$, we derive:
\begin{equation}
N(v_{\rm s}) \propto v_{\rm s}^{-1+1/(\alpha-1)}.
\end{equation}

\noindent Now,
\begin{equation}
\overline{v}(r_{0}) = \frac{\sum_{r=r_{0}} v_{i}}{N(r_{0})} =
\frac{ r_{0}^{1/\alpha-1} N(r_{0})} { N(r_{0})},
\end{equation}
resulting in
\begin{equation}
\overline{v_{\rm s}} \propto r_{\rm s}^{1-1/\alpha}.
\end{equation}
Substitution of the values for $\gamma$ and $\alpha$, results in:
\begin{eqnarray}
N(r_{\rm s}) \propto r_{\rm s}^{2/3}, \\
N(v_{\rm s}) \propto v_{\rm s}^{-7/2}, \\
\overline{v_{\rm s}} \propto r_{\rm s}^{-2/3}.
\end{eqnarray}

\item{\bf Single burst, luminosity spectrum.} If all shells were
created at the same instant and the input mechanical luminosity
has a power-law distribution then initial conditions are:
\begin{eqnarray}
\Phi(L_{\rm s}) \propto L_{\rm s}^{-\beta_{\rm ob}} r_{\rm s}
\propto L_{\rm s}^{\gamma} t_{0}^{\alpha} \propto L_{\rm
s}^{\gamma}.
\end{eqnarray}
Starting with:
\begin{equation}
N(r_{s}) dr_{s} = N(L) dL = L_{\rm s}^{-\beta_{\rm ob}} dL.
\end{equation}
Substituting $dr_{s}/dL \propto L_{\rm s}^{\gamma -1}$, we derive:
\begin{equation}
N(r_{\rm s}) \propto r_{\rm s}^{(1-\beta_{\rm ob})/\gamma -1}.
\end{equation}
Similarly, $v_{\rm s} \propto L_{\rm s}^{\gamma}$, resulting in:
\begin{equation}
N(v_{\rm s}) \propto v_{\rm s}^{(1-\beta_{\rm ob})/\gamma -1}.
\end{equation}
Now,
\begin{equation}
\overline{v}(r_{0}) = \frac{\sum_{r=r_{0}} v_{i}}{N(r_{0})} =
\frac{ v_{0} N(r_{0})} { N(r_{0})} \propto r_{0},
\end{equation}
resulting in
\begin{equation}
\overline{v_{\rm s}} \propto r_{\rm s}.
\end{equation}
Substituting values for $\gamma$ and $\alpha$, we get to:
\begin{eqnarray}
N(r_{\rm s}) \propto r_{\rm s}^{4-5\beta_{\rm ob}}, \\
N(v_{\rm s}) \propto v_{\rm s}^{4-5\beta_{\rm ob}}, \\
\overline{v_{\rm s}} \propto r_{\rm s}.
\end{eqnarray}
Expectedly, eq.19 is in agreement with the result of Oey and
Clarke (1997), for shells dominated by growing shells (shells
larger than 1300pc, according to their analysis).

\item {\bf Continuous creation, luminosity spectrum.} The last
case considers continuous shell creation together with the
power-law distribution with initial conditions given by:
\begin{eqnarray}
\Phi(L_{\rm s}) \propto L_{\rm s}^{-\beta_{\rm ob}} \frac{dN}{dt}
= const. = \psi .
\end{eqnarray}
At one particular time, $t=t_{0}$:
\begin{equation}
N(r_{s}) \partial r_{s} = N(L) \partial L.
\end{equation}
Substituting $L \propto r_{s}^{1/\gamma}/t_{0}^{\alpha/\gamma}$ we
get to:
\begin{equation}
N(r_{s}) \propto r_{s}^{\frac{1-\gamma-\beta}{\gamma}}
t_{0}^{\frac{\alpha(\beta-1)}{\gamma}}.
\end{equation}
Now, total number of shells can be expressed with:
\begin{equation}
N(r_{s}) = \int_{0}^{t=t_{max}}
r_{s}^{\frac{1-\gamma-\beta}{\gamma}}
t^{\frac{\alpha(\beta-1)}{\gamma}} dt, or
\end{equation}
\begin{equation}
N(r_{s}) = r_{s}^{\frac{1-\gamma-\beta}{\gamma}}
t_{max}^{\frac{\alpha(\beta-1)}{\gamma}+1}.
\end{equation}
Shells expand in the ISM until they reach $v_{\rm s} = v_{\rm
ISM}$, and at that stage $t_{max}=r_{s}/v_{\rm ISM}$. Substituting
this into the equation (25), we arrive at:
\begin{equation}
N(r_{s}) \propto r_{s}^{\frac{(1-\beta)(1-\alpha)}{\gamma}}.
\end{equation}
And following the same reasoning we derive:
\begin{eqnarray}
N(v_{s}) \propto v_{s}^{\frac{1}{\alpha-1} -1} \\
\overline{v_{\rm s}} \propto r_{\rm s}^{\frac{\alpha-1}{\alpha}}.
\end{eqnarray}
This corresponds to:
\begin{eqnarray}
N(r_{\rm s}) \propto r_{\rm s}^{2(1-\beta_{\rm ob})} \\
N(v_{\rm s}) \propto v_{\rm s}^{-7/2}, and \\
\overline{v_{\rm s}}(r_{\rm s}) \propto r_{\rm s}^{-2/3}.
\end{eqnarray}

\end{enumerate}

We now compare these predictions with the functions inferred from
observations (Section 3.3). Obviously, this comparison should be
carried out for Group I shells that do appear to have young
populations in them. However, the conclusions do not alter if all
shells are considered, as Group II shells have very similar
properties. According to the results described in Section 3, for
all shells, $\overline{v_{\rm s}}(r_{\rm s}) \propto r_{\rm
s}^{0.71\pm0.03}$. This result rules out cases (i) and (iii)
above. Also, the slopes for $N(r_{\rm s})$ and $N(v_{\rm s})$ were
found to be very similar (cf Table 3), which, again is only
satisfied by  case (ii), indicating that at least the majority of
the observed shells were created more or less in a single burst.
This is also depicted on Figure 7, where we have overlaid on the
$v_{\rm s}$ vs $r_{\rm s}$ data plane, lines of equal expansion
age. It is obvious that most of the shells have dynamical ages
close to 5 Myr (see also Paper I).

Since for case (ii), $\alpha_{\rm r_{\rm s}}=\alpha_{\rm v_{\rm
s}}$, we can now use the observed mean slope $\alpha=(\alpha_{\rm
r_{\rm s}}+\alpha_{\rm v_{\rm s}})/2=-2.19\pm0.23$, to estimate
the slope $\beta_{\rm ob}$ of the mechanical luminosity function
for the SMC: $4-5\beta_{\rm ob} = -2.19\pm0.23$, therefore,
$\beta_{\rm ob} = 1.24\pm0.05$. The slope of the HII LF estimated
from observations is $a_{\rm obs}=2\pm0.5$ (Kennicutt, Edgar \&
Hodge 1989). Therefore, the condition $\beta_{\rm ob}\leq a$
mentioned in Section 4.1 is satisfied.

To conclude, we have found that the shell radius and expansion
velocity distributions of the SMC shells and supershells  are
consistent with the shells having been formed in a single burst
and with an input mechanical luminosity spectrum $L_{\rm
s}^{-\beta_{\rm ob}}$, with  $\beta_{\rm ob} = 1.24\pm0.05$, which
is of the order of but smaller than the slope of the HII
luminosity function, as it would be expected.  Only growing shells
were considered.

 The same conclusions would be reached, if the distributions of Group
 II shells were considered instead of Group I. Actually, for Group II shells,
$\overline{v_{\rm s}}(r_{\rm s}) \propto r_{\rm s}^{0.87\pm0.08}$,
with an exponent even closer to the case (ii) prediction, than
Group I shells (which have a slope of $0.68\pm0.03$. This fact is
very difficult to understand within the framework of the standard
model. One would have to assume that Group II shells are actually
older (hence no young stars can be seen anymore within their
radius), but they have somehow been re-accelerated. However, one
would have to invoke a re-acceleration mechanism that would apply
to all Group II shells in the same way and at the same time, to
account for the tightness of the $\overline{v_{\rm s}}$ vs $r_{\rm
s}$ relation and its similarity to the Group I relation.

\subsection{Can Group II shells be GRB remnants ?}

\label{s:grb-stats}

We investigate here the possibility that Group II (empty) shells
are old remnants of GRBs. As mentioned in Section 1, recent
studies suggest that the GRBs could leave shell-like remnants in
the ISM (Efremov et al. 1998; Loeb
\& Perna 1998; Perna, Raymond \& Loeb 2000). As it is not expected
that the GRB remnants would be associated with young stellar
objects this could be one possible explanation for Group II
shells. In turn, finding GRB remnants is very important as it
would help to constrain types of environments where GRBs occur and
hence the GRB formation mechanisms.

The underlying source, connected to a GRB event, that produces the
initial explosion, releasing a large amount of kinetic energy into
the surrounding medium, is still unknown.
The most common theory is based on the collapse of a single super-massive star
(Paczynski 1998), so called `collapsar' scenario.
An alternative scenario involves a coalescence
of two compact objects, either two neutron stars in a binary
system or a neutron star and a black hole (Eichler et al. 1989).
As massive stars have short lives it is expected that their
remnants should be found in dense environments. On the other hand,
merging neutron stars are old objects and their remnants are
expected in intermediate to low-density environments (Perna,
Raymond \& Loeb 2000). It is believed that most of the GRBs are
beamed. The most recent results by Bloom, Frail, \& Kulkarni (2003)
suggest an extremely narrow range for the typical amount of energy that
GRBs deposit in the ISM: $8\times10^{50}$ -- $5\times10^{51}$ ergs,
after correction for the beaming effect.

GRB remnants display a characteristic double-shell morphology
during the intermediate age of $<$5 kyr (Ayal \& Piran 2001) which
distinguishes them from the OB shell remnants (resulting from
stellar winds and multiple supernovae explosions). At later stages
of their evolution GRB shells are very similar to typical OB
shells. Perna, Raymond \& Loeb (2000) suggested several spectral
and/or abundance signatures for their identification. The only
small difference between OB shells and GRB shells at the later
stages of evolution is in the nature of their growth: while most
of the growth of OB shells occurs due to the energy conservation
resulting from a continuous fueling by supernovae bursts, being
described by the Sedov-Taylor phase, GRB shells expand mostly as a
result of momentum conservation, which is described as the
snowplough phase. Hence, the expansion of GRB shells can be
expressed in a similar way to the expansion of OB shells
(discussed in Section 4.1).

\begin{figure}
\leavevmode  \epsfxsize=8cm \epsffile{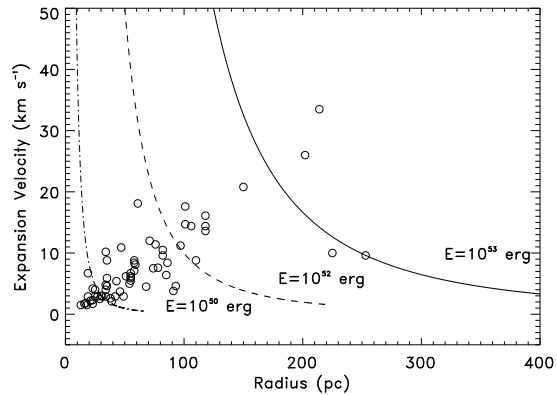}
\caption{\label{f:L} Comparison of observations of empty shells
(open circles) with predictions of the GRB model for the expansion
velocity versus shell radius relation.}
\end{figure}

Based on equations for the swept-up shell once it enters the
snowplough phase given by Cioffi, McKee, \& Bertschinger (1988),
the growth of a GRB-created shell in an uniform ISM can be
described by: $r_{\rm s} \propto L_{\rm s}^{\gamma} t^{\alpha}$
and $v \propto L_{\rm s}^{\gamma} t^{\alpha -1}$, where
$\gamma=0.22$ and $\alpha=0.30$.  On the observed $r_{\rm s}$ vs
$v_{\rm s}$ relation for Group II shells, we overlaid three equal
energy curves for GRB-created shells (Figure 8). Only shells with
energies in the range of $8\times10^{50}$ -- $5\times10^{51}$ ergs
can be considered as candidates for GRB-created shells,as
mentioned earlier. The diagram in Fig. 8 shows that there are
about 15--20 such shells. Their mean age is $(5\pm2)$ Myr. The
rest of the Group II shells have sizes and expansion velocities
too small or too large to be explained by GRB explosions. Now, if
all of the 15-20 shells were caused by GRBs, one would infer a GRB
rate of 3-4 Myr$^{-1}$, which is consistent with the predicted
frequency of GRBs in any galaxy by Schmidt (1999), for a beaming
factor of $\sim$0.03. Of course the value inferred here should be
treated as a lower limit, because the sample of empty shells
suffers from severe incompleteness, as such shells cannot be
identified in dense regions due to chance line-ups. Note also the
inhomogeneity in the spatial distribution of Group II shells in
the SMC (Fig.1).

We compare now the estimated GRB rate in the SMC, from the
potential GRB remnants, with the predictions based on the
``collapsar'' scenario. The estimated rate of Type II supernovae
in the SMC is $(500\pm300)$ per Myr (Crawford et al. 2000).
MacFadyen \& Woosley (1999) showed that collapsars are only a
small fraction of all supernovae, approximately 1\%. Hence, the
collapsar rate in the SMC can be estimated as $(5\pm3)$ per Myr.
This agrees well with the previously estimated (lower limit) GRB
rate. However, being associated with deaths of the most massive
stars, collapsars should be found in the regions with recent
active star-formation, while the great majority of the "empty"
shells are found in low-density environments with no young
populations. Therefore, although the statistics are encouraging,
it is unlikely that the shells in question have their origin in
collapsars.

Looking into the alternative ``merger'' scenario, the coalescence
rate of neutron stars in our Galaxy was estimated to be 17 per Myr
(Phinney 1991). Scaling this number by the ratio of potentially
observable pulsars in the Galaxy and in the SMC, which is equal to
0.03, gives an estimate of the coalescence rate of neutron stars
in the SMC of approximately 1 per Myr. This is lower than the
estimated rate, which is in any case a lower limit. However,
because the empty shells in question lie in an old stellar
environment, it cannot be ruled out that at least some of them,
about 5 in number if we adopt the neutron star coalescence rate
given above, could be the result of a merger GRB.

In conclusion, the mechanical luminosities of about 15--20 Group
II shells are consistent with the shells being GRB remnants, thus
 suggesting a GRB rate of $\geq$4 per Myr in the SMC, which
is consistent with  an estimate of the number of GRBs from the
collapse of super-massive stars. However, the latter would require
active recent star formation in the neighborhood of the shells,
which is not the case. On the other hand, GRBs caused by neutron
star coalescence occur in old stellar populations and could
account for about 5 of the Group II shells.

\subsection{Turbulence}

As mentioned in Section 1, the non-linear evolution of a
self-gravitating disk can also lead to formation of shells, that
are not related to a  central energy source (Wada \& Norman 1999).
As Elmegreen (1997) also suggested, HI `bubbles' (or shells) could
result naturally from the turbulent nature of the  interstellar
medium, i.e.  it is possible that most of the structure of the ISM
is the result of natural gaps and holes in the fractal gas
distribution caused by turbulence.
Even if turbulence alone cannot produce the required expansion
velocities, it is possible that supernovae, stellar winds and
ionizing radiation fill these gaps with ionized gas and provide
additional internal pressure.
There is
no modelling to date of the relation between expansion velocity
and size of shells formed by turbulence. But one might expect
(Elmegreen 2000, private communication) that the expansion
velocity would be  approximately proportional to the square root
of the shell radius ($v_{\rm s} \propto r_{\rm s}^{0.5}$), because
that is about what is seen for turbulence in general, i.e., for
the scaling of  clouds and collections of clouds. Observationally
we found that all shells observed follow the relation $v_{\rm s}
\propto r_{\rm s}^{0.71\pm0.03}$, which is close to the previous
qualitative statement. If most shells are indeed caused by
turbulence, that would explain why there is a relatively weak
correlation with OB associations and why the properties of ``empty"
and ``non-empty" shells are not very different in most cases.

Recently, Dib \& Burkert (2004) have investigated whether the
observed shells in the ISM of Holmberg II could be
produced by turbulence. Numerical simulations were performed for
different types of turbulent driving while taking into account
thermal and gravitational instabilities.
It appears that turbulence driven on large scales,
rather than supernova explosions, can reproduce
observed shells and holes in Holmberg II located in regions
without any detected stellar activity.
Future progress, as well as detailed comparison with
observations, in this direction is eagerly awaited.

\subsection{The role of the environment}
As was mentioned in the Introduction, regardless of the origin of
a shell, its evolution can be also affected by external factors:
radiation pressure from field stars (which is important for
supergiant shells), the density of the ambient ISM and its
inhomogeneities, interactions between shells, globular cluster
passage, etc. It is not clear, however, how one could
quantitatively identify observational signatures of such
processes, in order to proceed to a quantitative verification of
their action and relative importance. We described in Section
3.2.1 the difference in the behavior of inner and outer shells (of
either Group I or II) which could be accounted for by the
difference in the ambient density. We also noted that there is an
indication (Section 3.2) that larger shells follow a shallower
linear correlation on the $\log v_{\rm s}$ -- $\log r_{\rm s}$
plane than smaller shells. This could be the effect of age (with
larger shells being older), combined with environmental factors.
This latter statement however, is purely conjectural at this stage
and has to be further investigated in the future.


\section{Summary and Conclusions}
\label{s:summary}

There are 509 HI shell structures in the SMC, all apparently very
young, with dynamical ages of a few Myr. We have found that 59 of
these shells are empty, in the sense that they have no young
stellar objects associated with them. It is shown in Section 3
that on the whole properties of empty and non-empty shells are
very similar. Besides several individual cases, there is nothing
significantly different in shell properties -other than selection
biases- that could point to a different origin for these two
groups of shells. The reason could be a similar origin for all
shells in the SMC and/or a great importance that environmental
effects play in shaping shell characteristics especially at later
stages of evolution.

The shell radius and expansion velocity distribution functions are
consistent with the standard model, if all shells were created in
a single burst and the input mechanical luminosity had a power-law
distribution. This would indicate a burst of star formation at a
particular epoch (about 5 Myr ago). This interpretation however
cannot explain why the 59 shells with no young stellar
counterparts show almost exactly the same behavior as shells with
OB associations within their radius.

Sizes and expansion velocities for about 15-20 empty shells are
consistent with the expected properties for GRB remnants. These
shells suggest a GRB rate of  $>$4 per Myr in the SMC, which is
consistent with the predicted frequency of GRBs by Schmidt (1999)
for a beaming factor of $\sim$0.03, as well as with an estimate of
the number of GRBs from a collapse of super-massive stars.
However, the latter would require active recent star formation in
the neighborhood of the shells, which is not the case. The other
possibility is merger GRBs which requires an old stellar
environment. However, the predicted frequency of such events is
too low to account for more than $\simeq$ 5 of the 15-20 empty
shells in question. The rest of the empty shells have sizes and
expansion velocities too low or too high to be explained by GRB
explosions. So, on the whole, only a small number of the empty
shells may have been formed by GRBs.

We have also searched for morphological signatures that would
indicate a possible collision with a high velocity cloud for
"empty" shells. No significant features were found.

Most of the high luminosity empty shells appear to be associated
with a chimney-like feature in the NW outer regions of the SMC.

A comparison of the properties of shells lying in the inner and
outer regions of the SMC indicated that the density of the
environment within which a shell is expanding  might be
significantly affecting the evolution of a shell.

Therefore, none of the mechanisms described in the introduction
and examined here, can fully account for the properties of all of
the shells. Turbulence is a promising mechanism for the initiation
of shells, but detailed comparison with the observations was not
possible at this stage, due to lack of detailed models.

\bf {Acknowledgements} S. Stanimirovic was partially supported by
NSF grants AST-0097417 and AST-9981308, during this research. The
authors would also like to thank B. Elmegreen, Ch. Goudis, V.
Kalogera and R. Perna for useful discussions.


\begin{thebibliography}{99}
\bibitem{} Ayal S., Piran T., 2001, ApJ, 555, 23
\bibitem{} Azzopardi M., Breysacher J., 1979, A\&A, 75, 120
\bibitem{} Bica E.D., Schmitt H.R., 1995, A\&AS, 101, 41
\bibitem{} Bloom J.S., Frail D.A., \& Kulkarni S.R., 2003, ApJ, 594, 674
\bibitem{} Brinks E., Bajaja E., 1986, A\&A, 169, 14
\bibitem{} Bureau M., Carignan C., 2002, AJ, 123, 1316
\bibitem{} Cioffi D.F., McKee C.F., Bertschinger E., 1988, ApJ, 334, 252
\bibitem{} Crawford F., Gaensler B.M., Kapsi V.M., Manchester R.N., Camilo F.,
           Lyne A.G., Pivovaroff M.J., 2001, ApJ, 554, 152
\bibitem{} Dib S., Burkert A., 2004, ApJ, submitted (astro-ph/0402593)
\bibitem{} Eichler D., Livio M., Piran T., Schramm D.N., 1989, Nature, 340, 126
\bibitem[\protect\citename{Efremov, }1999]{Efremov99}Efremov, Y.~N.,
Ehlerova S., Palous J., 1999, ApJ, 350, 457
\bibitem[\protect\citename{Efremov et~al., }1998]{Efremov98}Efremov Y.~N.,
           Elmegreen B.~G., Hodge P.~W., 1998, ApJ, 501, 163L
\bibitem{} Elmegreen B.G., Chiang W.-H., 1982, ApJ, 253, 666
\bibitem{} Elmegreen B.G., 1997, ApJ, 477, 196
\bibitem{} Elmegreen B.G., Hunter, D., 2000, ApJ, 540, 814
\bibitem{} Filipovic M.D., Haynes R.F., White G.L., Jones P.A., 1998,
           A\&AS, 130, 421
\bibitem{} Gardiner L.T., Hatzidimitriou D., 1992, MNRAS, 257, 195
\bibitem{} Harris J. \& Zaritsky D., 2004, AJ, 127, 1531
\bibitem[\protect\citename{Heiles, Reach \& Koo, }1996]{Heiles96}
Heiles C., Reach W.~T., Koo B.-C., 1996, ApJ, 466, 191
\bibitem{} Heiles, C. 1984, ApJSuppl, 55, 585
\bibitem[\protect\citename{Heyer {et~al.}, }1996]{Heyer96}
Heyer M.~H., Brunt C., Snell R.~L., Howh J., Schloerb F.~P.,
Carpenter J.~M., Normandeau M., Taylor A.~R., Dewdney P.~E., Cao
Y., Terebey S., Beichman C.~A., 1996, ApJ, 464, L175
\bibitem{} Kennicutt R.C., Edgar B.K., Hodge P.W., 1989, ApJ, 337, 761
\bibitem{} Kim S., Dopita M.A., Staveley-Smith L., \& Bessell M.S.,
1999, AJ, 118, 2797
\bibitem{} Kontizas E., Morgan D.H., Kontizas M., Dapergolas A., 1988,
A\&A, 201, 208
\bibitem[\protect\citename{Loeb \& Perna, }1998]{Loeb98}
Loeb A., Perna R. 1998, ApJ, 503, 35L
\bibitem{} MacFadyen A.I., Woosley S.E., 1999, ApJ, 524, 262
\bibitem[\protect\citename{MacLow \& McCray, }1988]{MacLow88}
         MacLow M., McCray R., 1988, ApJ, 324, 776
\bibitem{} Maragoudaki F., Kontizas M., Morgan D.H., Kontizas E.,
           Dapergolas A., Livanou E., 2001, A\&A, 379, 864
\bibitem{} Mathewson D.S., Ford V.L., Visvanathan N., 1987, ApJ, 333, 617
\bibitem{} Maurice E., Bouchet P., Martin N., 1989, A\&AS, 78, 445
\bibitem{} McClure-Griffiths N.M., Dickey J.M., Gaensler B.M., Green A.J., Haynes
R.F., Wieringa M.H., 2000, AJ, 119, 2828
\bibitem{} McCray R., Kafatos M., 1987, ApJ, 317, 190
\bibitem{} Morgan D.H., Vassiliadis E., Dopita M.A., 1991, MNRAS, 251, 51
\bibitem{} Muller E., Staveley-Smith L., Zealey W., Stanimirovic
S., MNRAS, 339, 105
\bibitem[\protect\citename{Oey \& Clarke, }1997]{Oey97}
Oey M.~S., Clarke, C.~J., 1997, MNRAS, 289, 570
\bibitem{} Oey M.~S., Clarke, C.~J., 1999, In Proceedings of the
2nd Guillermo Haro Conference ``Interstellar Turbulence'', eds.
J. Franco and A. Carraminana, Cambridge U. Press, p. 112
\bibitem[\protect\citename{Paczy\'nski, }1998]{Paczynski98}
Paczy\'{n}ski B., 1998, ApJ, 494, 45L
\bibitem{}Perna R., Raymond J., 2000, ApJ, 539, 706
\bibitem{}Perna R., Raymond J., Loeb A., 2000, ApJ, 533, 658
\bibitem{}Phinney E.S., 1991, ApJ, 380, L17
\bibitem{} Puche D., Westpfahl D., Brinks E., Roy J.R., 1992, AJ, 103, 1841
\bibitem[\protect\citename{Rand \& Stone, }1996]{Rand96}
Rand R.~J., Stone J.~M., 1996, AJ, 111, 190
\bibitem{} Rhode K.L., Salzer J.J., Westfahl D.J., Radice L.A., 1999, AJ, 118, 323
\bibitem{} Savage A., Waldron l.D., Morgan D.H., Tritton S.B., Cannon R.D.,
Dawe J.A., Bruck M.T., Beard S.M., Palmer J.B., 1985a, The UK
Schmidt Telescope Objective Prisms: II. Illustrations of objective
Prism Spectra, Royal Observatory, Edinburgh
\bibitem{} Savage A., Waldron l.D., Fretwell M., Morgan D.H., Tritton S.B.,
Cannon R.D., Bruck M.T., Beard S.M., Palmer J.B., 1985b, The UK
Schmidt Telescope Objective Prisms: III. Illustrations of
objective Prism Spectra, Royal Observatory, Edinburgh
\bibitem{} Schmidt M., 1999, ApJ, 523L, 117
\bibitem{} Shostak G.S. \& Skillman E.D., 1989, A\&A, 214, 33
\bibitem{} Stanimirovic S., 1999, PhD Thesis, University of Western Sydney Nepean
\bibitem{} Stanimirovic S., Staveley-Smith L., Dickey J.M., Sault
R.J.,  Snowden S.L., 1999, MNRAS, 302, 417
\bibitem{} Staveley-Smith L., Sault R.J., Hatzidimitriou D., Kesteven M.J.,
McConnell D., 1997, MNRAS, 289, 225
\bibitem{} Stewart S.G., Walter F., 2000, AJ, 120, 1794
\bibitem{} Taylor A.~R., Irwin J.~A., Matthews H.~E., Heyer M.~H., 1999,
ApJ, 513, 339
\bibitem[\protect\citename{Tenorio-Tagle, }1981]{Tenorio-Tagle81}
Tenorio-Tagle G., 1981, A\&A, 94, 338
\bibitem[\protect\citename{Tenorio-Tagle {et~al.}, }1986]{Tenorio-Tagle86}
Tenorio-Tagle G., Bodenheimer P., Rozyczka M., Franco J., 1986,
A\&A, 170, 107
\bibitem{} Tenorio-Tagle G., Bodenheimer P., 1988, ARAA, 26, 145
\bibitem[\protect\citename{Tomisaka \& Ikeuchi, }1987]{Tomisaka87}
Tomisaka K., Ikeuchi S., 1987, PASJ, 38, 697
\bibitem{} van~der Hulst T., 1996, in The Minnesota Lectures on
Extragalactic Neutral Hydrogen ASP Conference Series, E.D.
Skillman, ed., 106, 47
\bibitem[\protect\citename{van~der Hulst, \& Sancisi, }1988]{vanderHulst881}
van~der Hulst T., Sancisi R., 1988, AJ, 95, 1354
\bibitem{} Wada K., Norman C.A., 1999, ApJ, 516L, 13
\bibitem{} Wada K., Spaans M., Kim S., 2000, ApJ, 540, 797
\bibitem{} Wallin J.F., Higdon J.L., Staveley-Smith L., 1996, ApJ, 459, 555
\bibitem{} Walter F. \& Brinks E., 1999, AJ, 118, 273
\bibitem{} Wang Q., Wu X., 1992, ApJSuppl, 78, 391
\bibitem[\protect\citename{Weaver et~al., }1977]{Weaver77}
Weaver R., McCray R., Castor J., Shapiro P., Moore R., 1977, ApJ,
218, 377
\bibitem{}Wilcots E.M. \& Miller B.W., 1998, AJ, 116, 2363
\bibitem{}Zaritsky D., Thompson, I.B., Grebel E.K., Massey P., 2002, AJ, 123, 855
\end{thebibliography}
\end{document}